\documentclass[11pt,a4paper]{article}
\usepackage{geometry} 
\usepackage{amsmath}
\usepackage{amsthm}
\usepackage{amssymb}
\usepackage{amsfonts}
\usepackage[latin1]{inputenc}

\usepackage{cite}

\def\be{\begin{equation}}
\def\ee{\end{equation}}
\def\bea{\begin{eqnarray}}
\def\eea{\end{eqnarray}}

\def\1{\'{\i}}                           
 
\def\conm#1#2{\left[ #1,#2 \right]}  
\def\pois#1#2{\left\{ #1,#2 \right\}}  

\begin{document}


\begin{center}
\ 

\bigskip

{\LARGE 
{\bf Curved momentum spaces from quantum groups 

\smallskip

with cosmological constant}}

\bigskip
\bigskip

{\sc \'A. Ballesteros$^1$, G. Gubitosi$^2$, I. Guti\'errez-Sagredo$^1$, F.J. Herranz$^1$}

{$^1$ Departamento de F\1sica, Universidad de Burgos, 
E-09001 Burgos, Spain}

{$^2$  Theoretical Physics, Blackett Laboratory,
Imperial College, London SW7 2AZ, United Kingdom
}
 
e-mail: {angelb@ubu.es, g.gubitosi@imperial.ac.uk, igsagredo@ubu.es, fjherranz@ubu.es}

\end{center}


\begin{abstract}
We  bring the concept that quantum symmetries describe theories with nontrivial momentum space properties one step further, looking at quantum symmetries of  spacetime in presence of a nonvanishing cosmological constant $\Lambda$. In particular, the momentum space associated to the $\kappa$-deformation of the de Sitter algebra in (1+1) and (2+1) dimensions is explicitly constructed as a dual Poisson-Lie group manifold parametrized by $\Lambda$. Such momentum space includes both the momenta associated to spacetime translations and the `hyperbolic' momenta associated to boost transformations, and has the geometry of (half of) a de Sitter manifold. Known results for the momentum space of the $\kappa$-Poincar\'e algebra are smoothly recovered in the limit $\Lambda\to 0$, where hyperbolic momenta decouple from translational momenta. The approach here presented is general and can be applied to other quantum deformations of kinematical symmetries, including (3+1)-dimensional ones.
\end{abstract}

\section{Introduction}

Recent  developments in quantum gravity research have revived and given new substance to the long-forgotten idea that momentum space should have a nontrivial geometry,  an intuition originally due to Max Born \cite{BornReciprocity}. 
After more than a decade since Deformed Special Relativity (DSR) was first proposed \cite{AmelinoCamelia:2000mn, AmelinoCamelia:2000ge}, it is now understood that a nontrivial geometry of momentum space is a general feature of DSR theories \cite{KowalskiGlikman:2003we, KowalskiGlikman:2002ft, Raetzel:2010je, AmelinoCamelia:2011bm, AmelinoCamelia:2011pe}. This is intimately related with the presence of the Planck energy as a second relativistic invariant (besides the speed of light), that can play the role of a curvature scale  of the momentum manifold \cite{Kowalski-Glikman:2013rxa}.
Nontrivial properties of momentum space  emerge also in (2+1)-dimensional quantum gravity, where explicit computations show that  the effective description of quantum gravity coupled to point particles is given by a theory with curved momentum space and noncommutative spacetime coordinates \cite{Matschull:1997du, Freidel:2003sp, Meusburger:2003ta, Freidel:2005me}. Of more direct interest for the results we are going to present here are  models of noncommutative geometry, where the space of momenta that are dual to the noncommutative spacetime coordinates  is curved \cite{Majid:1999tc, AmelinoCamelia:1999pm, KowalskiGlikman:2002jr, Gubitosi:2013rna}. 

Besides finding  increasing theoretical support, Planck-scale modifications of the geometry of momentum space are  extremely relevant from a phenomenological point of view. In fact, features due to curvature of momentum space are dual to those that in  general relativity are ascribed to curvature of spacetime:  in the same way as spacetime curvature induces redshift of energy, curvature of momentum space induces a dual redshift, that is,  an energy-dependent correction to the time of flight of free particles \cite{Amelino-Camelia:2013uya}.  Such effects open up a much needed observational window for Planck-scale physics, since they are testable with astrophysical  observations \cite{AmelinoCamelia:2008qg}.

Despite the recent significant theoretical and phenomenological progress just discussed, an important ingredient which is necessary to connect the properties of momentum space to observations is still missing. In fact, all of the models mentioned above are essentially deformations of special relativity: even though spacetime might be nontrivial (e.g. spacetime coordinates might not commute), still it has vanishing curvature. This is clearly a phenomenological shortcoming, since the most promising observations involve propagation of particles over cosmological distances, for which spacetime curvature cannot be neglected \cite{AmelinoCamelia:2009pg}. 
In the past few years several proposals  aimed at extending relativistic models with curved momentum space were put forward in order to include nonvanishing spacetime curvature. 
The first concrete approach \cite{AmelinoCamelia:2012it} focussed on constructing an extension of the Poincar\'e algebra that includes both the Planck scale and a (constant) spacetime curvature scale as relativistic invariants. The resulting algebra can be seen as a DSR version of the de Sitter (hereafter dS) algebra of symmetries, but the associated coalgebra was not investigated. Other proposals focussed on developing a unifying  description of the whole  phase space of free particles moving on a curved spacetime with deformed local Poincar\'e symmetries \cite{Barcaroli:2015xda, Barcaroli:2016yrl,  Barcaroli:2017gvg, Amelino-Camelia:2014rga, Cianfrani:2014fia}. The general understanding coming from these  approaches is that when both momentum space and spacetime have nonvanishing curvature they become so  intertwined that it is not possible to give a neat geometrical description of the properties of momentum space on its own.

In this work we show that this is not necessarily the case. Indeed, we are able to explicitly construct the curved momentum space generated by quantum-deformed spacetime symmetries in presence of a nonvanishing cosmological constant. We achieve this result by enlarging the momentum space so that it is not only the manifold of momenta associated to translations on spacetime, but it also includes the `hyperbolic' momenta associated to the boost transformations and the angular momenta associated to rotations. Within this construction we can also show that in the vanishing cosmological constant limit  the  Lorentz sector is not needed because it decouples from the energy-momentum sector, thus recovering previous results in the literature.

While we would like to argue that our results are general, we use the setting of Hopf algebras to  present an explicit derivation. Hopf algebras have proved to be a very useful mathematical framework to model DSR effects. The most studied example is the $\kappa$-Poincar\'e Hopf algebra \cite{Majid:1994cy, Lukierski:1992dt, Lukierski:1993wxa}, the investigation of which provided inspiration and more precise understanding of several  features of DSR models. For example, it can be explicitly shown  that  the manifold of momenta associated to the $\kappa$-Poincar\'e translation generators is a (portion of a) dS manifold, whose curvature is determined by the quantum deformation scale $\kappa$ \cite{KowalskiGlikman:2004tz, Gubitosi:2013rna} and whose metric determines the free particle dispersion relation that is indeed compatible with the $\kappa$-Poincar\'e symmetries, thus showing that the phenomenology associated to the $\kappa$-Poincar\'e algebra fits very naturally within the framework of relative locality \cite{Gubitosi:2013rna, AmelinoCamelia:2011nt}.

Here we present a generalization of all these results by working with the $\kappa$-deformation of the dS algebra (see~\cite{Lukierski:1991ff, Lukierski:1991pn, ck2, ck3, BHBruno, starodutsev, BHMNplb2017}). The name is due to the fact that in the limit of vanishing cosmological constant $\Lambda$ one recovers the $\kappa$-Poincar\'e algebra, while in the limit of vanishing quantum deformation parameter $z=1/\kappa$ one recovers the algebra of symmetries of the dS spacetime. It is worth noticing that it was exactly using  this Hopf algebra that the first pioneering investigations concerning the interplay between spacetime and momentum space curvature were undertaken \cite{Marciano:2010gq}. 

The Poisson version of the $\kappa$-dS Hopf algebra in (1+1) and in (2+1) dimensions is defined in section \ref{sec:kdS}, where it is shown that  the main  differences with respect to the corresponding $\kappa$-Poincar\'e structures fully arise in the (2+1) setting: whilst in the vanishing cosmological constant limit the translation generators $\{P_0,P_1,P_2\}$ close a Hopf subalgebra, this is no longer the case for the $\kappa$-dS algebra, since the cosmological constant mixes the translation and Lorentz sectors within both the coproduct map and the deformed Casimir function. 
Thus, for nonvanishing $\Lambda$ it seems natural to consider an enlarged momentum space including also the dual coordinates to the Lorentz generators. This idea allows us to construct the curved (generalized) momentum manifold in the nonvanishing cosmological constant setting as the full dual Poisson-Lie group manifold, whose explicit construction can be achieved through the Poisson version of the `quantum duality principle' (see~\cite{Dri, STS, GC, dualJPA} and references therein). 

The $\kappa$-dS dual Poisson-Lie groups are explicitly constructed in section \ref{sec:kdSdual}. In (1+1) dimensions the dual group coordinates are those associated to both the spacetime translations and boosts, and a certain linear action of the dual group on the origin of momentum space generates (half of) a (2+1)-dimensional dS manifold $M_{dS_3}$, spanned by the orbit of the group  passing through the origin. In this case, the fact that boosts have the same role in the momentum space as translation generators can be understood since their coproducts have the same formal structure. 
In (2+1) dimensions one spatial rotation comes into play and the structure of the $\kappa$-dS Hopf algebra is apparently much more involved. Nevertheless, the construction of the full dual Poisson-Lie group $G^\ast_\Lambda$ gives the clue for the full geometrical description of the associated momentum space.  The dual Lie algebra and its associated Poisson-Lie group are explicitly constructed  in section \ref{kdS3ddual}, and the corresponding linear action on the enlarged momentum space can be defined in such a way that the dual rotation generates the isotropy subgroup of the origin of the momentum space. As a consequence, we find  that  a (4+1)-dimensional  space of momenta associated to translations and boosts arises as a dual group orbit passing through the origin, and such a space again has the geometry of (half of) a dS manifold $M_{dS_5}$. 
Moreover, in the vanishing cosmological constant limit, the Lorentz sector completely decouples both in the dispersion relation and in the coproduct, thus recovering the well-known $\kappa$-Poincar\'e momentum space.
The paper ends with a concluding section in which the applicability of the method here presented to the construction of the $\kappa$-AdS momentum space is shown, and the keystones for solving the corresponding (3+1)-dimensional problem are presented.


\section{The $\kappa$-dS Poisson-Hopf algebra}\label{sec:kdS}

Let us start by reviewing the structural properties of the $\kappa$-deformation of the (1+1) and (2+1) dS algebra, which will be presented by considering the cosmological constant $\Lambda>0$ as an explicit parameter whose $\Lambda\to 0$ limit provides automatically the expressions for the $\kappa$-Poincar\'e algebra. In this way, the specific features of the construction leading to the $\kappa$-Poincar\'e momentum space will become transparent, and the proposed path to its nonvanishing cosmological constant generalization will arise in a natural way. 

In the subsection on the (1+1)-dimensional case we just briefly present the essential formulas, postponing a more in-depth discussion of the relevant features of the $\kappa$-dS algebra to the following subsection focussing on the (2+1)-dimensional case.

\subsection{The (1+1) $\kappa$-dS algebra}

The (undeformed) Poisson-Hopf dS algebra in (1+1) dimensions is defined by the brackets
\be
\pois{K}{P_0}=P_1,
\qquad
\pois{K}{P_1}= P_0 ,
\qquad
\pois{P_0}{P_1}=-\Lambda\,K,
\label{ds11}
\ee
where $K$ is the generator of boost transformations, $P_0$ and $P_1$ are the time and space translation generators and the (undeformed) coproduct is given by $\Delta_0(X)=X\otimes 1 + 1\otimes X$, with $X\in\{K,P_0,P_1\}$.
The Poisson version of the (1+1) $\kappa$-dS quantum algebra~\cite{ck2} is a Hopf algebra deformation of~\eqref{ds11}, given by
\be
\pois{K}{P_0}=P_1,
\qquad
\pois{K}{P_1}=\frac{\sinh\left( z P_0 \right)}{z} ,
\qquad
\pois{P_0}{P_1}=- \Lambda\,K,
\label{qds}
\ee
with deformed coproduct map
\begin{eqnarray}
\Delta(P_0)&=&P_0 \otimes 1 + 1 \otimes P_0 , \nonumber\\
\Delta(P_1)&=& P_1\otimes e^{\frac{z}{2} P_0} +  e^{-\frac{z}{2} P_0} \otimes P_1, \label{coin2}\\
\Delta(K)&=& K\otimes e^{\frac{z}{2} P_0} +  e^{-\frac{z}{2} P_0} \otimes K.
\nonumber
\end{eqnarray}  
The quantum deformation parameter is $z=1/\kappa$ and the deformed Casimir function for~\eqref{qds} is
\be
C_z=\left(\frac{\sinh\left( z P_0 /2 \right)}{z/2}\right)^2 - P_1^2 + \Lambda\, K^2.
\ee
The so-called bicrossproduct-type basis~\cite{Majid:1994cy} for this algebra is given through the nonlinear change
\be
P_0 \to P_0,
\qquad
P_1 \to e^{\frac{z}{2} P_0} \,P_1,
\qquad
K \to e^{\frac{z}{2} P_0} \,K,
\label{newcoord}
\ee
so that the algebra becomes
\be
\pois{K}{P_0}=P_1,
\qquad
\pois{K}{P_1}=\frac{ 1-\exp(-2 z P_0)}{2z}-\frac{z}{2}\,(P_1^2 - \Lambda\, K^2) ,
\qquad
\pois{P_0}{P_1}=- \Lambda\,K,
\label{kdsa11}
\ee
with associated coproduct map 
\begin{eqnarray}
\Delta(P_0)&=&P_0 \otimes 1 + 1 \otimes P_0, \nonumber\\
\Delta(P_1)&=& P_1\otimes 1 +  e^{-{z}P_0} \otimes P_1,\label{coin2}\\
\Delta(K)&=& K\otimes 1 +  e^{-{z} P_0} \otimes K.
\nonumber
\end{eqnarray}  
In this basis, the deformed Casimir reads
\be
C_z=\left(\frac{\sinh\left( z P_0 /2 \right)}{z/2}\right)^2 - e^{{z} P_0} (P_1^2 - \Lambda K^2).
\label{kcas11}
\ee
We point out that for $\Lambda= 0$ (the $\kappa$-Poincar\'e case), the momentum sector given by $P_0$ and $P_1$ generates an Abelian Hopf subalgebra, and the $\Lambda=0$ Casimir function provides the well-known (1+1) $\kappa$-Poincar\'e deformed dispersion relation (see e.g.~\cite{Gubitosi:2013rna}). Note also that the coproduct~\eqref{coin2} does not depend on $\Lambda$, although this property will not hold in higher dimensions.


\subsection{The (2+1) $\kappa$-dS algebra}

In (2+1) dimensions, the Poisson-Lie brackets of  the (undeformed)
dS algebra  take the form
\be
\begin{array}{lll} 
\pois{J}{P_i}=   \epsilon_{ij}P_j , &\qquad
\pois{J}{K_i}=   \epsilon_{ij}K_j , &\qquad  \pois{J}{P_0}= 0  , \\[2pt]
\pois{P_i}{K_j}=-\delta_{ij}P_0 ,&\qquad \pois{P_0}{K_i}=-P_i ,&\qquad
\pois{K_1}{K_2}= -J   , \\[2pt]
\pois{P_0}{P_i}=-\Lambda\, K_i ,&\qquad \pois{P_1}{P_2}= \Lambda\, J  ,
\end{array}
\label{ba} 
\ee
where $i,j=1,2$, and  $\epsilon_{ij}$ is a skew-symmetric tensor with $\epsilon_{12}=1$ (note that for negative values of $\Lambda$,  this bracket  defines the AdS Poisson-Lie algebra). The two quadratic Casimir functions for~\eqref{ba}  are 
\be
{\cal C}=P_0^2-\mathbf{P}^2- \Lambda(J^2-\mathbf{K}^2), \qquad
{\cal W}=-JP_0+K_1P_2-K_2P_1  ,
\label{bc}
\ee
where $\mathbf{P}^2=P_1^2+P_2^2$ and $\mathbf{K}^2=K_1^2+K_2^2$. Recall that $\cal C$ comes from to  the Killing--Cartan form and is related to the energy of a point particle, while $\cal W$ is the Pauli--Lubanski vector. The undeformed Hopf algebra structure is given by $\Delta_0$.

The (2+1) $\kappa$-dS Poisson-Hopf algebra
in the bicrossproduct basis is the Hopf algebra deformation with parameter $z=1/\kappa$ given by~\cite{ck3, BHBruno, starodutsev}
\be
\begin{array}{lll} 
\pois{J}{P_0}=   0    ,
&  \pois{J}{P_1}=   P_2 ,
& \pois{J}{P_2}=  -P_1
  , \\[2pt]
\pois{J}{K_1}=     K_2    ,
&   \pois{J}{K_2}=     -K_1   ,
&   \pois{K_1}{K_2}=  -\frac{  \sin(2 z \sqrt{\Lambda} J)}{2z \sqrt{\Lambda} }   , \\[4pt]
   \pois{P_0}{P_1}=  - \Lambda\,K_1 ,
&    \pois{P_0}{P_2}=   - \Lambda\,K_2  ,
&  \pois{P_1}{P_2}=   \Lambda\, \frac{  \sin(2 z \sqrt{ \Lambda} J)}{2z \sqrt{\Lambda} } , 
\end{array}
\nonumber
\ee
\vskip-0.75cm
\bea
&& \pois{K_1}{P_0}=     P_1  ,\qquad\qquad\qquad\qquad\;\;\;
    \pois{K_2}{P_0}=     P_2,\label{p21}\\
&& \pois{P_2}{K_1}=     z \left(P_1 P_2 - \Lambda K_1 K_2\right)    \qquad \pois{P_1}{K_2}=    z \left( P_1 P_2- \Lambda K_1 K_2  \right)      ,
\nonumber\\
&& \pois{K_1}{P_1}=      \frac{1}{2z} \left(  \cos(2z\sqrt{\Lambda} J) - e^{-2zP_0}    \right)  +\frac{z}{2} \left( P_2^2-P_1^2\right)  - \frac{z \Lambda}{2}  \left( K_2^2-K_1^2 \right)  ,\nonumber\\
&& \pois{K_2}{P_2}=   \frac{1}{2z} \left( \cos(2z \sqrt{ \Lambda}  J) - e^{-2zP_0} \right) +\frac{z}{2} \left(P_1^2 -P_2^2\right)-\frac{z\Lambda}{2} \left(  K_1^2 -K_2^2 \right)        ,\nonumber
\eea
and with deformed coproduct map 
\bea
\Delta ( P_0 ) \!\!\!&=&\!\!\!  P_0 \otimes 1+1 \otimes P_0   ,\qquad \Delta  ( J  ) =  J \otimes 1 +1 \otimes J , \nonumber\\
\Delta ( P_1 )  \!\!\!&=&\!\!\!  P_1\otimes   \cos (z\sqrt{ \Lambda} J) +e^{-z P_0}  \otimes P_1 
+  \Lambda\, K_2 \otimes    \frac{ \sin (z\sqrt{\Lambda} J) } {\sqrt{ \Lambda}}  , \nonumber \\
\Delta ( P_2 ) \!\!\!&=&\!\!\!  P_2\otimes  \cos( z\sqrt{\Lambda} J) +e^{-z P_0}  \otimes P_2
- \Lambda\, K_1 \otimes   \frac{ \sin (z\sqrt{\Lambda} J) } {\sqrt{ \Lambda}}  , \label{bc} \\
\Delta ( K_1 ) \!\!\!&=&\!\!\!  K_1\otimes     \cos( z\sqrt{ \Lambda} J) +e^{-z P_0}  \otimes K_1 +  P_2 \otimes       \frac{ \sin (z\sqrt{\Lambda} J) } {\sqrt{\Lambda}}   ,  \nonumber \\
\Delta ( K_2 ) \!\!\!&=&\!\!\!  K_2\otimes    \cos (z\sqrt{\Lambda} J) +e^{-z P_0}  \otimes K_2 -  P_1 \otimes      \frac{ \sin (z\sqrt{\Lambda} J) } {\sqrt{\Lambda}}  ,
\nonumber
\eea 
which explicitly depends on the cosmological constant $\Lambda$.
The deformed Casimir function for this Poisson-Hopf algebra reads
\be
{\cal C}_z=\frac 2{z^2}\left[ \cosh (zP_0)\cos(z\sqrt{ \Lambda} J)-1 \right]
-e^{zP_0} \left( \mathbf{P}^2 - \Lambda \,\mathbf{K}^2 \right)\,\cos(z\,\sqrt{\Lambda}\,J)
-2\, \Lambda \,e^{zP_0}\,\frac{\sin(z \sqrt{\Lambda}J)}{\sqrt{ \Lambda}}\,R_3,
\label{cz21}
\ee
with $R_3=\epsilon_{3bc} K_b P_c$.  
Note that the projection to the $\kappa$-dS algebra in (1+1) dimensions is obtained by setting to zero the generators $\{P_2,K_2,J\}$.

The (2+1) $\kappa$-Poincar\'e Hopf algebra is smoothly recovered in the $\Lambda\to 0$ limit and in this `flat' case the momentum sector $\{P_0,P_1,P_2\}$ generates an Abelian Hopf subalgebra with coproducts
\bea
\Delta ( P_0 ) \!\!\!&=&\!\!\!  P_0 \otimes 1+1 \otimes P_0, \nonumber\\
\Delta ( P_1 )  \!\!\!&=&\!\!\!  P_1\otimes   1+e^{-z P_0}  \otimes P_1 , \label{kp11} \\
\Delta ( P_2 ) \!\!\!&=&\!\!\!  P_2\otimes 1 +e^{-z P_0}  \otimes P_2.
\nonumber
\eea
Such a nonlinear superposition law for momenta is the essential footprint of a curved momentum space, which can be explicitly constructed by following the procedure presented in~\cite{Kowalski-Glikman:2013rxa}. Essentially, the $\kappa$-Poincar\'e momentum space is a three-dimensional manifold generated by the action on a certain ambient space of the three-dimensional dual Lie group $G^\ast$ whose Lie algebra $g^\ast$,
\be
\conm{X^0}{X^1}=-z\,X^1,
\qquad
\conm{X^0}{X^2}=-z\,X^2,
\qquad
\conm{X^1}{X^2}=0, \label{km}
\ee
is defined as the dual of the skew-symmetric part of the first order deformation in $z$ of the coproducts~\eqref{kp11}. The Lie algebra~\eqref{km} is the so-called (2+1) $\kappa$-Minkowski noncommutative spacetime~\cite{kMinkowski, Majid:1994cy}. 
Moreover, when $\Lambda=0$ the deformed Casimir function
 \bea
{\cal C}_z\!\!\!&=&\!\!\! \frac 2{z^2}\left[ \cosh (zP_0)-1 \right]
-e^{zP_0} \,(P_1^2+P_2^2),
\eea
provides the $\kappa$-Poincar\'e deformed dispersion relation in (2+1) dimensions.
The same construction can be straightforwardly generalized to the (3+1) $\kappa$-Poincar\'e algebra (see~\cite{Kowalski-Glikman:2013rxa} and references therein). 

The main obstruction to a similar construction when $\Lambda\neq 0$ is readily seen by inspection of eq.~\eqref{bc}. In fact, in the $\kappa$-dS case the momentum sector $\{P_0,P_1,P_2\}$ is no longer a Hopf subalgebra, since the coproduct of spatial momenta includes all the  generators $\{J,K_1,K_2\}$ of the Lorentz sector (note that this is not the case in (1+1) dimensions, where the coproduct does not depend on $\Lambda$). Moreover, the deformed Casimir ${\cal C}_z$ contains the Lorentz generators as well, and this feature is also present in the (1+1) case (see eq.~\eqref{kcas11}). These two observations hold true also in the (3+1) $\kappa$-dS Poisson-Hopf algebra that has been explicitly presented for the first time in~\cite{BHMNplb2017}. 

We already mentioned that Hopf-algebraic deformations of spacetime symmetries can be endowed with a phenomenological interpretation. Specifically, the Casimir ${\cal C}_z$ of the algebra determines the dispersion relation of free particles, while the coproduct of the translation generators determines the rules of conservation of energy and spatial momentum in interactions \cite{Gubitosi:2013rna}. Therefore, when $\Lambda\neq 0$ we can say that both the conservation rules in interactions and the deformed dispersion relation involve an enlarged set of `momenta', including also the angular momentum and the `hyperbolic' momenta corresponding, respectively, to the rotation and to boost transformations (hyperbolic rotations). In this framework, it seems natural to propose that when $\Lambda\neq 0$ the (curved) momentum space is defined by an enlarged space parametrized by the six coordinates that are dual to the generators of the full quantum algebra. Nevertheless, a simple inspection at the coproducts~\eqref{bc} shows that the role of the $J$ generator is somewhat different from that of $K_{1}$ and $K_2$, since the latter have coproducts which are formally equivalent to those of $P_1$ and $P_2$.  All these aspects will have a clear interpretation once the explicit construction of the $\kappa$-dS momentum space is performed in the following section. 


\section{Momentum space for the $\kappa$-dS Poisson-Hopf algebra}\label{sec:kdSdual}

As anticipated above, in this section the momentum space for the
$\kappa$-dS Poisson algebra with nonvanishing cosmological constant will be constructed as the full dual Poisson-Lie group $G^\ast_\Lambda$, whose Lie algebra $g^\ast_\Lambda$ is provided by the dual of the cocommutator map $\delta$ generated by the coproduct of all the $\kappa$-dS generators in the bicrossproduct basis, including the Lorentz sector. This construction will be firstly illustrated in (1+1) dimensions. While this case is simpler, it does not allow to appreciate the richness of structure characterising higher-dimensional models. The consistency and geometric features of our approach will be made fully explicit in the second subsection, where we demonstrate the full construction for the (2+1)-dimensional case.

\subsection{The (1+1) case}

The cocommutator map for the full $\kappa$-dS algebra can be read from the skew-symmetric part of the first-order deformation in $z$ of the coproduct~\eqref{coin2}, namely
\be
\delta(P_0)=0, \qquad\ 
\delta(P_1)= z\, P_1 \wedge P_0, \qquad
\delta(K)=z\, K \wedge P_0.
\label{delta}
\ee
If we denote by  $\{X^0,X^1,L\}$  the  generators dual to, respectively, $\{P_0,P_1,K\}$, the dual Lie algebra $g^\ast_\Lambda$ is  given by the Lie brackets
\be
\conm{X^0}{X^1}=-z\,X^1,
\qquad
\conm{X^0}{L}=-z\,L,
\qquad
\conm{X^1}{L}=0.
\label{kM}
\ee
A faithful representation $\rho$ of this Lie algebra for $\Lambda\neq 0$  is given by the $4 \times 4$ matrices
\be
\rho(X^0)=z \left( 
\begin{array}{cccc}
0 & 0 & 0 & 1\\
0 & 0 & 0 & 0\\
0 & 0 & 0 & 0\\
1 & 0 & 0 & 0
\end{array}
\right)  \quad 
\rho(X^1)=z\left( 
\begin{array}{cccc}
0 & 1 & 0 & 0\\
1 & 0 & 0 & 1\\
0 & 0 & 0 & 0\\
0 & -1 & 0 & 0
\end{array}
\right)  \quad 
\rho(L)=z \sqrt{\Lambda}\left( 
\begin{array}{cccc}
0 & 0 &1 & 0\\
0 & 0 & 0 & 0\\
1 & 0 & 0 & 1\\
0 & 0 & -1 & 0
\end{array}
\right).  \quad 
\label{repbook}
\ee
If we denote as $\{p_0,p_1,\chi \}$ the local group coordinates which are dual, respectively, to $\{X^0,X^1,L \}$, 
then the group element of the dual Lie group $G^\ast_\Lambda$ is given by:
\be
G^\ast_\Lambda=\exp \left(p_1 \rho(X^1) \right) \exp \left( \chi\rho(L) \right) \exp \left(p_0 \rho(X^0) \right).
\ee
A straightforward computation leads to the following explicit matrix
\be
G^\ast_\Lambda=
\left( 
\begin{array}{cccc}
\cosh(z p_0) \, +\frac{1}{2}\,e^{z\,p_0}\,z^2 (p_1^2 + \Lambda \chi^2)& z p_1 & z \sqrt{ \Lambda} \chi & \sinh(z p_0) \, +\frac{1}{2}\,e^{z\,p_0}\,z^2 (p_1^2 + \Lambda \chi^2)\\
e^{z\,p_0}\,z p_1& 1 & 0 & e^{z\,p_0}\,z p_1\\
e^{z\,p_0}\, z \sqrt{\Lambda}\chi & 0 & 1 & e^{z\,p_0}\, z \sqrt{\Lambda} \chi\\
\sinh(z p_0) \, -\frac{1}{2}\,e^{z\,p_0}\,z^2 (p_1^2 +\Lambda \chi^2) & -z p_{1} & -z \sqrt{\Lambda} \chi & \cosh(z p_0) \, -\frac{1}{2}\,e^{z\,p_0}\,z^2 (p_1^2 + \Lambda \chi^2)
\end{array}
\right).
\label{element}
\ee

The multiplication law for the group $G^\ast_\Lambda$ is  obtained by multipying two matrices of the form~\eqref{element}, and it can be written as a co-product (see~\cite{dualJPA}) in the form
\be
\Delta(p_0)=p_0 \otimes 1 + 1 \otimes p_0, \qquad
\Delta(p_1)=p_1\otimes 1 +  e^{-{z} p_0} \otimes p_1, \qquad
\Delta(\chi)= \chi\otimes 1 +  e^{-{z} p_0} \otimes \chi.
\label{codual}
\ee
As the quantum duality principle indicates, this coproduct is just the one~\eqref{coin2} for the $\kappa$-dS algebra once one identifies  the dual group coordinates and the generators of the $\kappa$-dS Poisson-Hopf algebra as follows:
\be
p_0\equiv P_0, \quad
p_1\equiv P_1, \quad
\chi\equiv K.
\label{id11}
\ee

Moreover, by following the technique presented in~\cite{dualJPA} it can be shown that the unique Poisson-Lie structure on $G^\ast_\Lambda$ that is compatible with the coproduct~\eqref{codual} and has the undeformed dS Lie algebra~\eqref{ds11} as its linearization is given by the Poisson brackets
\be
\pois{\chi}{p_0}=p_1,
\qquad
\pois{\chi}{p_1}=\frac{ 1-\exp(-2 z p_0)}{2z}-\frac{z}{2}\,(p_1^2  - \Lambda\, \chi^2) ,
\qquad
\pois{p_0}{p_1}=- \Lambda\,\chi,
\ee
which is exactly the $\kappa$-dS algebra~\eqref{kdsa11} under the identification~\eqref{id11}.
Evidently, the Casimir function for this Poisson bracket is
\be
C_z=\left(\frac{\sinh\left( z p_0 /2 \right)}{z/2}\right)^2 - e^{{z} p_0} (p_1^2  - \Lambda \chi^2).
\label{kcas11dual}
\ee

In this way, the composition law for the momenta with $\kappa$-dS symmetry~\eqref{coin2} has been reobtained as the group law~\eqref{codual} for the coordinates of the dual Poisson-Lie group $G^\ast_\Lambda$, and the $\kappa$-dS Casimir function~\eqref{kcas11} can be interpreted as an on-shell relation~\eqref{kcas11dual} for these coordinates. 

We stress that the main novelty with respect to the $\kappa$-Poincar\'e case described in~\cite{Kowalski-Glikman:2013rxa} is the fact that the dual Lie group $G^\ast_\Lambda$ is now three-dimensional, and the momentum space associated to $\kappa$-dS is parametrized by the three coordinates $\{p_0,p_1,\chi \}$, and not only by the  momenta associated to spacetime translations. Moreover, both in the coproduct~\eqref{codual} and the Casimir function~\eqref{kcas11dual} the role of the parameters $\chi$ and $p_1$ turns out to be identical, which supports the role of the former as an additional `hyperbolic' momentum for quantum symmetries with nonvanishing cosmological constant.

An explicit geometric interpretation of this enlarged momentum space can be obtained along the same lines of~\cite{Kowalski-Glikman:2013rxa} by observing that the entries of the fourth column in $G^\ast_\Lambda$, given by
\begin{eqnarray}
S_0&=&\sinh(z p_0) \, +\frac{1}{2}\,e^{z\,p_0}\,z^2 (p_1^2 + \Lambda \chi^2), \nonumber\\
S_1&=& e^{z\,p_0}\,z\, p_1, \nonumber\\
S_2&=& e^{z\,p_0}\, z\, \sqrt{\Lambda}\, \chi, \\
S_3&=& \cosh(z p_0) \, -\frac{1}{2}\,e^{z\,p_0}\,z^2 (p_1^2 + \Lambda \chi^2), \nonumber
\nonumber
\end{eqnarray}  
satisfy the defining relation for the (2+1)-dimensional dS space,
\be
-S_0^2 + S_1^2 + S_2^2 + S_3^2 =1.
\ee
 Moreover, if we consider a linear action of the Lie group $G^\ast_\Lambda$ onto a four-dimensional ambient Minkowski space with coordinates $(S_0,S_1,S_2,S_3)$,  we have that
\be
G^\ast_\Lambda\cdot
(0,0,0,1)^T=(S_0,S_1,S_2,S_3)^T,
\ee
which means that the (2+1)-dimensional dS space is generated through $G^\ast_\Lambda$ as the orbit that passes through the point $(0,0,0,1)$ in the ambient space, corresponding to the origin of the (generalized) momentum space. Note that the orbit passing through the point $(0,0,0,\alpha)$, with $\alpha\neq 0$, would satisfy  $-S_0^2 + S_1^2 + S_2^2 + S_3^2 =\alpha^2$. 
Moreover, we have that the condition
\be
S_0+S_3=e^{z\,p_0}>0,
\ee 
is automatically obeyed, so that only half of the (2+1)-dimensional dS space is generated as an orbit of the free action of $G^\ast_\Lambda$, and we will denote this manifold as $M_{dS_3}$.  Finally, when $\Lambda=0$ the ambient coordinate $S_2$ vanishes, as well as the realization $\rho(L)$ of the dual of the boost generator, thus recovering the well-known  interpretation of the $\kappa$-Poincar\'e momentum space as (half of) a (1+1)-dimensional dS space, {\em i.e.}, $M_{dS_2}$.


\subsection{The (2+1) case}
\label{kdS3ddual}

The very same procedure described in the previous section can be applied to the construction of the momentum space associated to the (2+1) $\kappa$-dS Poisson-Hopf algebra. The skew symmmetrized first order in $z$ of the coproduct~\eqref{bc} is given by the cocommutator map
\begin{eqnarray}
&& \delta(P_0) =  \delta(J)=0 , \nonumber\\ 
&&  \delta (P_1)=   z (P_1\wedge P_0  + \Lambda \, K_2\wedge J  )  ,
\nonumber\\ 
&&
  \delta(P_2)=  z  (P_2\wedge P_0 - \Lambda\,  K_1 \wedge J)       ,
\label{cc}  \\
&&
  \delta(K_1)= z (K_1 \wedge P_0  + P_2 \wedge J)   ,
 \nonumber\\ 
&&
  \delta(K_2)=   z  (K_2  \wedge P_0 -P_1\wedge J)   .
  \nonumber
\end{eqnarray}
Denoting by $\{X^0,X^1,X^2,L^1,L^2,R\}$  the  generators dual to, respectively, $\{P_0,P_1,P_2,K_1,K_2,J\}$, the Lie brackets defining the Lie algebra $g^\ast$ of the dual Poisson-Lie group $G^\ast_\Lambda$ are
\be
\begin{array}{lll} 
[X^0, X^1]=-z \, X^1    ,
& \qquad [X^0, X^2]=-z \, X^2  , 
& \qquad [X^1, X^2]=0
  , \\[2pt]
[X^0, L^1]=-z\,  L^1    ,
&\qquad [X^0, L^2]=-z \, L^2  , 
&\qquad [L^1, L^2]=0  ,
 \\[2pt]
[R, X^2]=-z \, L^1    ,
 &\qquad[R, L^1]=z\,{\Lambda}\,  X^2  , 
&\qquad [L^1, X^2]=0   ,
 \\[2pt]
[R, X^1]=z \, L^2    ,
 &\qquad[R, L^2]=-z\,{\Lambda}\,  X^1  , 
 &\qquad[L^2, X^1]=0  ,
 \\[2pt]
[R, X^0]=0    ,
&\qquad [L^1, X^1]=0  , 
 &\qquad[L^2, X^2]=0  .
\end{array}
\label{lie21}
\ee
A (faithful) representation $\rho$ of this Lie algebra for $\Lambda\neq 0$  is given by the $6 \times 6$ matrices
\be
\rho(X^0)= z \left( 
\begin{array}{cccccc}
0 & 0 & 0 & 0 & 0 & 1\\
0 & 0 & 0 & 0 & 0 & 0\\
0 & 0 & 0 & 0 & 0 & 0\\
0 & 0 & 0 & 0 & 0 & 0\\
0 & 0 & 0 & 0 & 0 & 0\\
1 & 0 & 0 & 0 & 0 & 0
\end{array}
\right)  \quad 
\rho(X^1)= z \left( 
\begin{array}{cccccc}
0 & 1 & 0 & 0 & 0 & 0\\
1 & 0 & 0 & 0 & 0 & 1\\
0 & 0 & 0 & 0 & 0 & 0\\
0 & 0 & 0 & 0 & 0 & 0\\
0 & 0 & 0 & 0 & 0 & 0\\
0 & -1 & 0 & 0 & 0 & 0
\end{array}
\right) 
\nonumber
\ee
\be
\rho(X^2)= z \left( 
\begin{array}{cccccc}
0 & 0 & 1 & 0 & 0 & 0\\
0 & 0 & 0 & 0 & 0 & 0\\
1 & 0 & 0 & 0 & 0 & 1\\
0 & 0 & 0 & 0 & 0 & 0\\
0 & 0 & 0 & 0 & 0 & 0\\
0 & 0 & -1 & 0 & 0 & 0
\end{array}
\right)
 \quad 
\rho(L^1)= z \sqrt{ \Lambda} \left( 
\begin{array}{cccccc}
0 & 0 & 0 & 1 & 0 & 0\\
0 & 0 & 0 & 0 & 0 & 0\\
0 & 0 & 0 & 0 & 0 & 0\\
1 & 0 & 0 & 0 & 0 & 1\\
0 & 0 & 0 & 0 & 0 & 0\\
0 & 0 & 0 & -1 & 0 & 0
\end{array}
\right) 
\label{6drep}
\ee
\be
\rho(L^2)= z \sqrt{\Lambda} \left( 
\begin{array}{cccccc}
0 & 0 & 0 & 0 &1 & 0\\
0 & 0 & 0 & 0 & 0 & 0\\
0 & 0 & 0 & 0 & 0 & 0\\
0 & 0 & 0 & 0 & 0 & 0\\
1 & 0 & 0 & 0 & 0 & 1\\
0 & 0 & 0 & 0 & -1 & 0
\end{array}
\right)
\quad
\rho(R)= z \sqrt{ \Lambda} \left( 
\begin{array}{cccccc}
0 & 0 & 0 & 0 & 0 & 0\\
0 & 0 & 0 & 0 &-1 & 0\\
0 & 0 & 0 & 1 & 0 & 0\\
0 & 0 & -1 & 0 & 0 & 0\\
0 & 1 & 0 & 0 & 0 & 0\\
0 & 0 & 0 & 0 & 0 & 0
\end{array}
\right).
\nonumber
\ee
If we denote as $\{p_0,p_1,p_2,\chi_1,\chi_2,\theta \}$ the local group coordinates which are dual, respectively, to $\{X^0,X^1, X^2, L^1,L^{2}, R \}$,
then the  Lie group element $G^\ast_\Lambda$ can be written as
\be
G^\ast_\Lambda=\exp \left(\theta \rho(R)\right)\exp \left(p_1 \rho(X^1)\right)\exp \left(p_2 \rho(X^2) \right) \exp \left( \chi_1\rho(L^1) \right)
\exp \left( \chi_2\rho(L^2) \right) \exp \left(p_0 \rho(X^0) \right),
\label{ge21}
\ee
and its explicit expression can be straightforwardly computed, although we omit it here for the sake of brevity. By multiplying two of these generic group elements, the group law for $G^\ast_\Lambda$ can be directly derived and written as the following coproduct map for the six dual group coordinates:
\begin{eqnarray}
\Delta(p_0)&=&p_0 \otimes 1 + 1 \otimes p_0, \qquad
\Delta(\theta)=\theta \otimes 1 + 1 \otimes \theta, \nonumber\\
\Delta(p_1)&=& p_1\otimes \cos(z\,\sqrt{\Lambda}\, \theta) +  e^{-{z} p_0} \otimes p_1 +\Lambda\,\chi_2\otimes 
\frac{\sin(z\, \sqrt{\Lambda}\, \theta)}{\sqrt{\Lambda}},
\nonumber\\
\Delta(p_2)&=& p_2\otimes \cos(z\, \sqrt{ \Lambda}\, \theta) +  e^{-{z} p_0} \otimes p_2  - \Lambda\,\chi_1\otimes
\frac{\sin(z\, \sqrt{\Lambda}\, \theta)}{\sqrt{\Lambda}}, \label{codual21}\\
\Delta(\chi_1)&=& \chi_1\otimes \cos(z\, \sqrt{ \Lambda}\, \theta) +  e^{-{z} p_0} \otimes \chi_1  + p_2\otimes 
\frac{\sin(z\, \sqrt{\Lambda}\, \theta)}{\sqrt{\Lambda}},\nonumber\\
\Delta(\chi_2)&=& \chi_2\otimes \cos(z\, \sqrt{ \Lambda}\, \theta) +  e^{-{z} p_0} \otimes \chi_2  - p_1\otimes \frac{\sin(z\, \sqrt{\Lambda}\, \theta)}{\sqrt{\Lambda}}.
\nonumber
\end{eqnarray}  
Again, under the identification
\be
p_0\equiv P_0, \quad
p_1\equiv P_1, \quad
p_2\equiv P_2, \quad
\chi_1\equiv K_1, \quad
\chi_2\equiv K_2, \quad
\theta \equiv J,
\label{id21}
\ee
this is exactly the coproduct for the $\kappa$-dS Poisson-Hopf algebra given in~\eqref{bc}, and the unique Poisson-Lie structure on $G^\ast_\Lambda$ that is compatible with~\eqref{codual21} and has the undeformed dS Lie algebra (\ref{ba}) as its linearization is the deformed Poisson algebra given by~\eqref{p21}.

In order to provide a geometric interpretation of the six-dimensional generalized momentum space manifold, we proceed similarly to the (1+1) case and consider the action of $G^\ast_\Lambda$ onto an ambient space. The entries of the sixth column in the matrix realization~\eqref{ge21} are
\begin{eqnarray}
S_0&=& \sinh(z p_0) \, +\frac{1}{2}\,e^{z\,p_0}\,z^2 \left(p_1^2 + p_2^2 +\Lambda \left(\chi_1^2+ \chi_2^2\right)\right), \nonumber\\
S_1&=& e^{z\,p_0}z\,(\cos(z\,\sqrt{\Lambda}\,\theta)\, p_1- \sqrt{\Lambda}\,\sin(z\,\sqrt{\Lambda}\,\theta)\,\chi_2 ), \nonumber\\
S_2&=& e^{z\,p_0}z\,(\cos(z\,\sqrt{\Lambda}\,\theta)\, p_2 + \sqrt{\Lambda}\, \sin(z\,\sqrt{\Lambda}\,\theta)\,\chi_1 ), \nonumber\\
S_3&=& e^{z\,p_0}z\,(-\sin(z\,\sqrt{\Lambda}\,\theta)\, p_2+ \sqrt{\Lambda}\, \cos(z\,\sqrt{\Lambda}\,\theta)\,\chi_1 ), \label{cms21}\\
S_4&=& e^{z\,p_0}z\,(\sin(z\,\sqrt{\Lambda}\,\theta)\, p_1+ \sqrt{\Lambda}\,\cos(z\,\sqrt{\Lambda}\,\theta)\,\chi_2), \nonumber\\
S_5&=&\cosh(z p_0) \, -\frac{1}{2}\,e^{z\,p_0}\,z^2 \left(p_1^2 + p_2^2 +\Lambda \left(\chi_1^2+ \chi_2^2\right)\right),\nonumber\end{eqnarray}  
and satisfy the condition
\be
-S_0^2 + S_1^2 + S_2^2 + S_3^2 + S_4^2 + S_5^2 =1,
\label{qds21}
\ee
which is the defining relation for the (4+1)-dimensional dS space.  Therefore, by assuming that the space of generalized momenta is the group manifold for the dual group $G^\ast_\Lambda$, we can conclude that a linear action of the Lie group $G^\ast_\Lambda$ onto a six-dimensional ambient Minkowski space with coordinates $(S_0,S_1,S_2,S_3,S_4,S_5)$ allows us to obtain a (4+1) dS space as the orbit that passes through the point in the ambient space with coordinates  $(0,0,0,0,0,1)$, which is the origin of the (generalized) momentum space. Moreover, we have that $
S_0+S_5=e^{z\,p_0}>0,
$
so only half of the dS space is generated in this way, and we will denote this manifold as $M_{dS_5}$. Therefore, the (1+1) construction can be generalized to this (2+1) setting, although some distinctive features of the latter are worth to be stressed.

Firstly, given that in the (2+1) case one has six symmetry generators, one would naively expect that the generalized momentum space be a six dimensional manifold, given that in the (1+1) case the dimensionality of the manifold corresponds to the number of symmetry generators. Instead,  we demonstrated the emergence of a five-dimensional orbit under the action of $G^\ast_\Lambda$. The reason for this is the completely different role that the dual rotation $(R,\theta)$ plays with respect to the dual boosts $(L^i,\chi_i)$, both in the coproduct and in the action~\eqref{cms21}. In particular, it is immediate to check that the isotropy subgroup of the point $(0,0,0,0,0,1)$ is just the one given by $G_0^\ast=\exp \left(\theta \rho(R)\right)$. Therefore, the full momentum space for the $\kappa$-dS algebra in (2+1) dimensions is the six-dimensional manifold $M_{dS_5} \times S^1$, where the rotation coordinate $\theta$ is the one parametrizing $S^1$ while $(p_i, \chi_i)$ parametrize $M_{dS_5}$.

Secondly, under the identification~\eqref{id21} the deformed Casimir  is written as the following function on the generalized momentum space:
\be
{\cal C}_z = \frac 2{z^2}\left[ \cosh (zp_0)\cos(z\sqrt{\Lambda} \theta)-1 \right]
-e^{zp_0} \left( p_1^2 +p_2^2 - \Lambda ({\chi}_1^2 + \chi_2^2)\right) \cos(z\sqrt{\Lambda}  \theta)
-2\,\Lambda\,e^{zp_0}  \frac{\sin(z \sqrt{ \Lambda}\theta)}{\sqrt{ \Lambda}}R_3,
\label{czr}
\ee
which involves all the translation and Lorentz momenta. Nevertheless, if we specialize this function onto the five-dimensional orbit $M_{dS_5}$ by taking the $S^1$ coordinate $\theta=0$, we get
\be
{\cal C}_z = \frac 2{z^2}\left[ \cosh (zp_0)-1 \right]
-e^{zp_0} \left( p_1^2 +p_2^2 - \Lambda ({\chi}_1^2 + \chi_2^2) \right),
\label{czsr}
\ee
which is an on-shell relation that is just a higher dimensional generalization of the one obtained in the (1+1) $\kappa$-dS case, eq.~\eqref{kcas11dual}. In this way, the striking equivalence between the role played by the momenta associated to space translations and boosts is manifestly shown. 

Finally, the (2+1) $\kappa$-Poincar\'e construction is again straightforwardly recovered in the limit $ \Lambda\to 0$, where the action~\eqref{cms21} provides $S_3=S_4=0$ and the representation~\eqref{6drep} is only defined for $\{X^0,X^1,X^2\}$, thus giving rise to (half of) a (2+1) dS space as an orbit under the action of the corresponding three-dimensional dual group. Summarizing, in (2+1) dimensions the momentum space for $\kappa$-dS is found to be the six-dimensional manifold $M_{dS_5} \times S^1$, while its $\kappa$-Poincar\'e limit was known to be the three-dimensional one $M_{dS_3}$.


\section{Concluding remarks}

Deformed special relativity (DSR) theories are characterized by the presence of an energy scale that plays the role of a second relativistic invariant besides the speed of light. Such an energy scale allows the geometry of momentum space to be nontrivial, and in fact it is a general feature of DSR models that the manifold of momenta has nonzero curvature.

In this paper we have shown that the curved momentum space construction can be extended to cases where also a nonvanishing spacetime cosmological constant is present.
We explored in particular the momentum space of the $\kappa$-deformation of the dS algebra, called $\kappa$-dS, and we showed that one can construct a curved generalized-momentum space, that includes not only the momenta associated to spacetime translations but also the hyperbolic momenta  associated to boosts. The procedure is an adaptation of the one that was successfully used to show that the momentum space of the $\kappa$-Poincar\'e algebra has the geometry of (half of) a dS manifold and is generated by the orbits of the dual Poisson-Lie group. The construction here presented can be applied to any other Hopf algebra deformation of kinematical symmetries with nonvanishing $\Lambda$, although the orbit structure of the momentum space so obtained will indeed depend on the chosen quantum deformation.

The construction in (1+1) dimensions is quite straightforward once one realizes that the boosts play a very similar role to spatial translations in the structure of the algebra and coalgebra. We indeed found that the generalized-momentum manifold is a (2+1)-dimensional dS manifold, whose coordinates are the local group coordinates associated to spacetime translations and boosts.

In (2+1) dimensions matters are complicated by the presence of a rotation generator in the algebra, that significantly complicates its structure. However the rotation generator has a peculiar role in the structure of the algebra and coalgebra, while boosts still behave similarly to spatial translations. We were indeed able to construct the generalized momentum space of the (2+1) $\kappa$-dS algebra whose coordinates are the local group coordinates associated to spacetime translations and boosts, and we showed that this is half of a (4+1)-dimensional dS manifold, for which the dual rotation generator generates the isotropy subgroup of the origin.

It is worth mentioning that the formalism here presented, in which $\Lambda$ is considered as an explicit `classical' deformation parameter (and this fact is connected with the so-called `semidualization' approaches in (2+1) quantum gravity~\cite{MSsemid,OSsemid}), suggests the possibility of performing the same construction of the generalized momentum space for the $\kappa$-AdS (Anti de Sitter) algebra by taking $\Lambda<0$. It turns out that one can indeed work out fully the $\kappa$-AdS counterpart of the results described above. The main difference between the $\kappa$-dS and $\kappa$-AdS cases arises from the dual group representation~\eqref{6drep}, which has to be modified in the $\Lambda<0$ case in order to have a real representation of the corresponding dual Lie group $G^\ast_\Lambda$. The latter can be explicitly constructed and leads to an action on the point $(0,0,0,0,0,1)$ that generates the quadric
\be
-S_0^2 + S_1^2 + S_2^2 - S_3^2 - S_4^2 + S_5^2 =1,
\label{qads21}
\ee
which is no longer the $M_{dS_5}$ momentum space. Nevertheless, the $\Lambda\to 0$ limit of this action annihilates the $S_3$ and $S_4$ coordinates, thus giving rise to the same $\kappa$-Poincar\'e limit as the one previously obtained from the $\kappa$-dS algebra, as it should be.

While the point of this paper was clearly made limiting the analysis to lower-dimensional algebras, the application of the approach here presented to the construction of the momentum space for the (3+1)-dimensional $\kappa$-dS algebra seems to be feasible. In fact, the full $\kappa$-dS Poisson Hopf algebra in (3+1) dimensions has been recently presented in~\cite{BHMNplb2017}, and its corresponding momentum space should be obtained as a 10-dimensional dual Poisson-Lie group manifold by mimicking the procedure here presented.  In fact, by direct inspection of the expressions for the coproduct and cocommutator map in the (3+1) $\kappa$-dS algebra~\cite{BHMNplb2017}, the formal similiarity between  boosts and spatial momenta is again evident, while the three rotations are composed in a completely different way. This is work in progress and will be presented elsewhere, including the analysis of the momentum space for the (3+1) $\kappa$-AdS algebra. Also, it would be interesting to use this approach in order to construct the curved momentum spaces for other quantum deformations of kinematical symmetries with nonvanishing cosmological constant, like for instance the (2+1) (A)dS quantum group recently introduced in~\cite{BHMplb2} or other possible quantum (A)dS deformations (see~\cite{tallin, LukiBorowiec}).


\section*{Acknowledgments}

A.B., I.G-S and F.J.H. have been partially supported by Ministerio de Econom\'{i}a y Competitividad (MINECO, Spain) under grants MTM2013-43820-P and   MTM2016-79639-P (AEI/FEDER, UE), by Junta de Castilla y Le\'on (Spain) under grants BU278U14 and VA057U16 and by the Action MP1405 QSPACE from the European Cooperation in Science and Technology (COST). G.G. acknowledges support from the John Templeton Foundation.


\end{document}